\documentclass[sigconf,nonacm]{acmart}
\renewcommand\footnotetextcopyrightpermission[1]{}
\usepackage{pifont}   
\usepackage{xcolor}
\usepackage{graphicx}
\usepackage{amsmath}

\begin{document}

\title{SLED-IFV: Solver-Validated LLM-Guided Decomposition for Scalable Hardware Information-Flow Verification}

\settopmatter{authorsperrow=4}

\author{Liangtao Dai}
\affiliation{\institution{University of Virginia}\city{Charlottesville}\country{USA}}
\email{kvf4sf@virginia.edu}

\author{Yimin Gao}
\affiliation{\institution{University of Virginia}\city{Charlottesville}\country{USA}}
\email{yg9bq@virginia.edu}

\author{Melika Morsali}
\affiliation{\institution{University of Virginia}\city{Charlottesville}\country{USA}}
\email{qfc2zn@virginia.edu}

\author{Mircea Stan}
\affiliation{\institution{University of Virginia}\city{Charlottesville}\country{USA}}
\email{mircea@virginia.edu}

\renewcommand{\shortauthors}{Dai et al.}

\begin{abstract}
Formal hardware information-flow verification (IFV) provides strong guarantees against secret-dependent timing and control behavior, but often scales poorly on realistic RTL. We identify two recurring proof barriers in self-composed IFV: \emph{implementation complexity}, where proof-hard datapath logic dominates even though the property needs only a compact boundary relation, and \emph{relational inductive complexity}, where the proof depends on cross-copy public-control facts that the backend prover does not infer efficiently. To address them, we introduce two semantic proof decomposition forms: \emph{functional simplification}, which replaces a proof-hard RTL region with a validated over-approximate summary, and \emph{relational strengthening}, which exposes and proves the cross-copy relations needed for induction.
We further present \textbf{SLED-IFV}, a solver-validated LLM-guided flow that automates the selection of these forms and their concrete targets. Given a self-composed miter and an oracle-free decision sheet, the LLM proposes a decomposition, then materializes it into proof artifacts under controller checks. The controller compiles the checked artifacts into proof obligations, and the formal verification backend remains the sole authority for acceptance.
Across nine nontrivial benchmarks constructed from real RTL, SLED-IFV achieves up to \(603\times\) solver-only speedup and converts two 12-hour timeouts into completed proofs. The closed-loop flow produces verifier-accepted decompositions for all cases.
\end{abstract}



\keywords{Information-flow verification, formal verification, hardware security, property-directed reachability, large language models}

\maketitle

\section{Introduction}

Security bugs that escape into silicon are costly to patch and difficult to contain\cite{8835233,281330, 236236, 10.1145/3787109.3815316}. Modern SoCs therefore require signoff-level guarantees for hardware roots of trust, cryptographic accelerators, privilege logic, and constant-time execution. Hardware information-flow verification (IFV) provides such guarantees by proving \emph{non-interference}: changing a protected source must not affect a public observation.
A standard IFV encoding is \emph{self-composition}, in which two RTL copies execute with equal public inputs and potentially different protected sources while the verifier proves equality of the selected public observations~\cite{Barthe2004selfcomposition,Gleissenthall2021constranhwverification, Terauchi2005SelfComp}. This reduction turns non-interference into a safety property and enables IC3/PDR-style model checking~\cite{Bradley2011IC3,Een2011abcpdr}. In practice, however, such proofs are often substantially harder than ordinary functional checking because the backend must reason about two RTL executions simultaneously and infer the inductive facts that keep them aligned.

The central obstacle is not simply RTL size; the proof presented to the solver is often in the wrong form. A human verifier may recognize that a masked datapath can be summarized by a compact boundary law, or that two executions remain aligned because their public control states evolve in lockstep. A generic PDR engine sees only the duplicated transition relation and must recover both implementation-level facts and cross-copy relations from low-level RTL.

We find that hard RTL IFV proofs repeatedly expose two corresponding barriers. The first is \textbf{implementation complexity}, where proof-hard datapath logic dominates the cone of influence even though the proof requires only a compact boundary fact. Masked cryptographic logic is a representative example: schemes such as domain-oriented masking randomize internal shares~\cite{gross2016dom}, while the property may depend only on the corresponding unmasked value. The second is \textbf{relational inductive complexity}, where the difficulty lies in establishing cross-copy invariants such as equality of public counters, protocol phases, valid signals, or control states.

To address these barriers, we introduce two semantic proof decomposition forms. \emph{Functional simplification} separates proof-hard implementation detail from the residual IFV argument by replacing a candidate RTL region with an over-approximate summary whose boundary law is proved on the concrete implementation. \emph{Relational strengthening} exposes missing inductive structure by requiring PDR to prove the original IFV property together with candidate cross-copy relations on the concrete two-copy RTL. The two forms can be composed when both datapath and relational barriers occur in the same query.

Prior work improves hardware IFV scalability at several layers. Obligation-level methods reformulate properties or use structural analysis~\cite{Deutschmann2024UPEC-DIT,Deutschmann2025Fastpath}; information-flow tracking and static systems propagate security labels~\cite{hu2021hardwareift,zhang2015secverilog, ferraiuolo2017practical}; and backend techniques strengthen IC3/PDR through copy symmetry, equivalence predicates, or security-specific reasoning~\cite{tan2026secic3, dai2026guardedequivalencepredicatesscalable}. These techniques are complementary, but they largely operate within a selected verification formulation and leave the query-specific proof-decomposition space unexplored. This space is consequential because an unsuitable decomposition leaves the dominant proof bottleneck intact, whereas an appropriate one can turn an otherwise intractable IFV proof into tractable solver obligations. It is also difficult to navigate: the verifier must select not only among functional simplification, relational strengthening, and their composition, but also the concrete module boundary or cross-copy signals to target. These choices depend jointly on the source, observation, environment assumptions, RTL semantics, and candidate signal roles, and are therefore generally difficult to determine from RTL hierarchy or fixed structural rules alone.

We present \textbf{SLED-IFV}, a solver-validated, LLM-guided flow that explores this query-specific proof-decomposition space. Given a self-composed miter and an oracle-free decision sheet, the LLM first selects a functional, relational, or composed decomposition together with its concrete RTL or signal-level target, and then materializes the proposal into proof artifacts under controller-defined constraints. The controller checks the generated artifacts, compiles the valid ones into solver obligations, and leaves acceptance entirely to the formal verifier. To the best of our knowledge, SLED-IFV is the first framework to use an LLM to select and realize solver-validated proof decompositions for formal hardware IFV.

The LLM remains outside the trusted proof base. Generated contracts, summaries, auxiliary relations, and negative controls affect the result only after passing controller-side scope checks and conventional solver obligations, including local-law proofs, accelerated proofs, false-law canaries, and faithfulness checks. An incorrect proposal or artifact may waste solver time, but it cannot certify an incorrect result. Prior LLM-assisted hardware-verification methods primarily focus on general verification tasks, such as assertion generation and test generation~\cite{AssertLLM,zhang2025llm4dvusinglargelanguage}. In contrast, relatively little work has explored LLMs for formal hardware IFV. LLM-IFT~\cite{mashnoor2025llm} uses an LLM to infer information-flow violations directly from RTL, but does not couple its predictions with a formal proof backend. SLED-IFV instead places every LLM-generated decomposition and proof artifact inside a solver-validated flow, where the formal verifier remains the sole authority for acceptance.

We evaluate SLED-IFV on nine nontrivial RTL IFV cases from public designs and published masking components, including masked datapaths incorporating OpenTitan DOM components, Goldschmidt-based floating-point divider, HMAC/SHA logic, lowRISC Ibex, YosysHQ PicoRV32, and a Chipyard/BOOM-derived routing control~\cite{opentitan1,ibex_rtl,picorv32_rtl,chipyard, DIV_VLSITSA, DIV_ISCAS}. Using the same ABC-PDR backend~\cite{Een2011abcpdr}, SLED-IFV achieves up to 603$\times$ solver-only speedup and converts two 12-hour PDR timeouts into completed proofs. Ablation studies further show that functional simplification and relational strengthening address complementary proof barriers. 

This paper makes the following contributions:
\begin{itemize}
\item We identify two recurring barriers in RTL IFV--- \textbf{implementation complexity} and \textbf{relational inductive complexity}---and introduce two corresponding semantic proof decomposition forms: \emph{functional simplification} and \emph{relational strengthening}.
\item We present \textbf{SLED-IFV}, which uses an LLM to automate the query-specific selection of decomposition forms and concrete targets while retaining a fully solver-checked trust boundary.
\item We evaluate SLED-IFV on nine nontrivial RTL IFV cases, achieving up to 603$\times$ solver-only speedup, converting two 12-hour timeouts into completed proofs, and demonstrating that the two decomposition forms address complementary proof barriers.
\end{itemize}

\begin{figure}[t]
    \centering
    \includegraphics[width=0.78\linewidth]{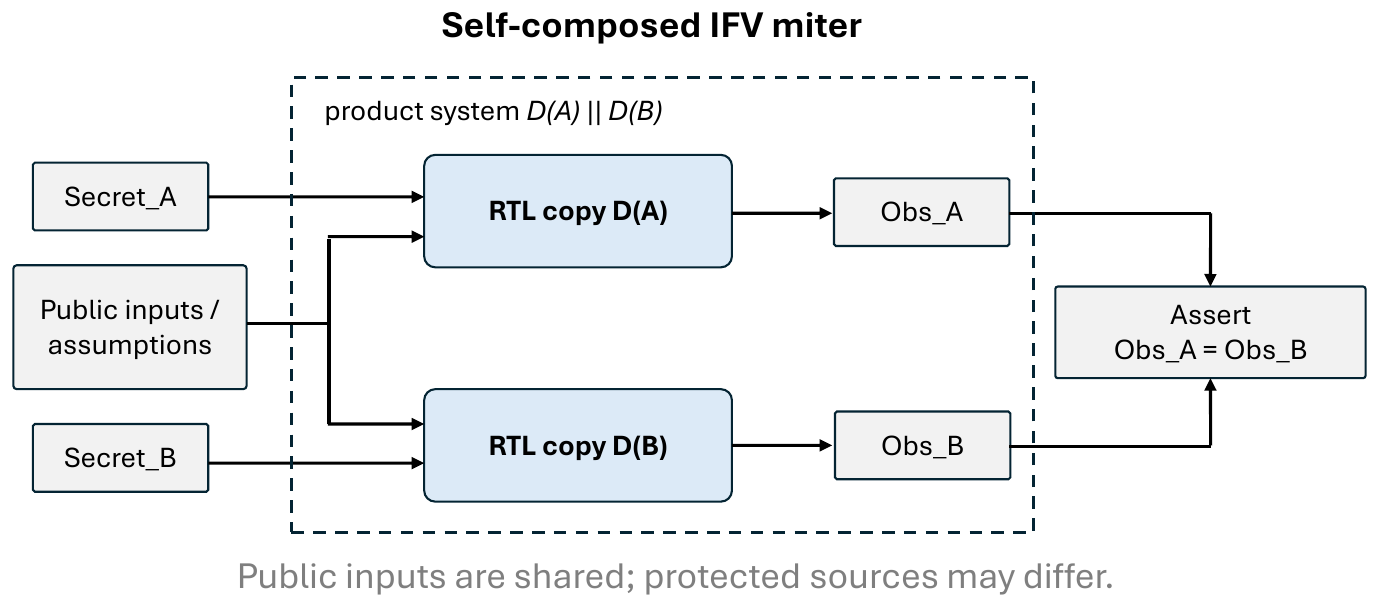}
    \caption{Self-composed IFV miter that reduces non-interference to a safety assertion over two RTL copies with shared public inputs and potentially different protected sources.}
    \label{fig:self-composed}
    \Description{Block diagram of a self-composed information-flow miter: two identical RTL copies receive equal public inputs while their protected sources may differ, and their public observations are compared through an equality assertion, reducing non-interference to a safety check over the product system.}
\end{figure}

\section{Background}
\label{sec:background}

Fig.~\ref{fig:self-composed} shows the standard self-composed IFV miter. Two RTL copies execute under the same public inputs and environment assumptions, while their protected sources may differ. The resulting observations are compared through a safety assertion, reducing non-interference to model checking over the product system.

Formally, we represent an RTL IFV task as
\[
    Q = (D,S,O,A),
\]
where \(D\) is the RTL design, \(S\) is the protected source, \(O\) is the public observation, and \(A\) denotes the public-input and environment assumptions. Self-composition instantiates two copies, \(D^{(0)}\) and \(D^{(1)}\), and defines the target property
\[
    P_{\mathrm{IFV}} \equiv O^{(0)} = O^{(1)}.
\]
Proving
\[
    D^{(0)} \parallel D^{(1)} \models P_{\mathrm{IFV}}
\]
establishes that the selected observation is independent of the protected source under \(A\).

We use IC3/PDR as the backend prover. On the product transition system in Fig.~\ref{fig:self-composed}, PDR establishes \(P_{\mathrm{IFV}}\) by constructing an inductive invariant over the combined state of the two RTL copies.

\begin{table}[t]
\centering
\caption{Representative IFV queries and their corresponding proof bottlenecks.}
\label{tab:motivation-cases}
\small
\setlength{\tabcolsep}{6pt}
\renewcommand{\arraystretch}{0.90}
\resizebox{\columnwidth}{!}{%
\begin{tabular}{@{}lllc@{}}
\toprule
\textbf{Case} & \textbf{Source $\rightarrow$ sink} & \textbf{Bottleneck} & \textbf{Form} \\
\midrule
Masked GF mult. &
\begin{tabular}[t]{@{}l@{}}
masking enc. \\
$\rightarrow$ unmasked product
\end{tabular} &
\begin{tabular}[t]{@{}l@{}}
masked nonlinear \\
datapath
\end{tabular} &
F \\

Ibex DMEM NI &
\begin{tabular}[t]{@{}l@{}}
DMEM word \\
$\rightarrow$ gated access trace
\end{tabular} &
\begin{tabular}[t]{@{}l@{}}
public-control \\
alignment
\end{tabular} &
R \\

Mini-cipher loop &
\begin{tabular}[t]{@{}l@{}}
masking enc. \\
$\rightarrow$ final unmasked state
\end{tabular} &
\begin{tabular}[t]{@{}l@{}}
datapath + loop \\
alignment
\end{tabular} &
F+R \\
\bottomrule
\end{tabular}}

\vspace{0.25em}
{\scriptsize F = functional simplification; R = relational strengthening.}
\vspace{-0.8em}
\end{table}
\vspace{-3mm}

\section{Motivation}
\label{sec:motivation}


Hard IFV queries can arise from qualitatively different proof bottlenecks. Depending on the source, observation, assumptions, and RTL semantics, the dominant difficulty may lie in proof-hard datapath logic, missing cross-copy inductive relations, or both. Table~\ref{tab:motivation-cases} illustrates these three representative patterns.

The masked GF multiplier is dominated by datapath implementation complexity. The two copies use different share encodings and fresh randomness, while the underlying unmasked operands are constrained to agree. The property observes the unmasked product, \(prod[0]\oplus prod[1]\). Although the concrete cone contains masked nonlinear share logic, the proof needs only the boundary unmasking relation. Functional simplification is therefore appropriate: the DOM implementation can be replaced by a summary once that boundary law has been proved on the concrete module. Relational strengthening provides little value here because the internal shares are not expected to match across copies.

The Ibex DMEM-access query exhibits the opposite behavior. The protected source is a data-memory word, and the sink is the gated memory-access trace, including request/write behavior and address-relevant control when a request is issued. Under a fixed constant-time program, the two executions should follow the same public control path. The main difficulty is therefore not a proof-hard datapath, but the absence of cross-copy lockstep facts needed for induction. Relational strengthening addresses this barrier by making selected public-control equalities explicit and proving them together with the original IFV property.

The mini-cipher round loop combines both effects. Independent masks and randomness create a proof-hard masked S-box, while the iterative round schedule introduces a separate control-alignment requirement. Functional simplification removes the masked datapath barrier but leaves the loop-alignment problem; relational strengthening exposes the round-control relation but leaves the nonlinear masked implementation intact. Neither form is sufficient in isolation, so the two must be composed.

These examples show that different IFV queries expose different dominant proof bottlenecks, and therefore require different decomposition strategies. The choice involves both the decomposition form and its concrete target: a module boundary for functional simplification, a set of public-control signals for relational strengthening, or both. SLED-IFV automates this bottleneck-aware selection, while accepting a proposal only after the corresponding solver obligations succeed.

\begin{figure}
    \centering
    \includegraphics[width=0.9\linewidth]{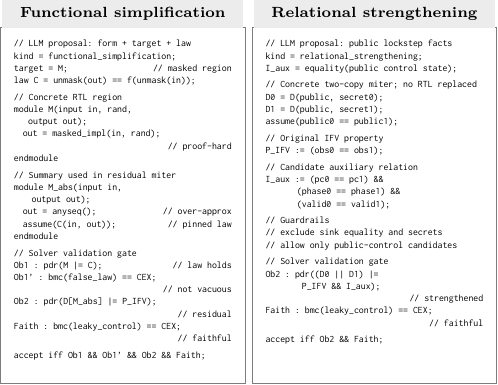}
    \caption{Code-level solver gates for SLED-IFV. Functional simplification validates a local summary before proving the residual miter; relational strengthening proves \(P_{\mathrm{IFV}} \wedge I_{\mathrm{aux}}\) on the concrete two-copy RTL.}
    \label{fig:decomposition-forms}
    \Description{Illustration of the two solver-side gates in SLED-IFV: functional simplification first proves a local boundary law on the concrete module and then proves the residual miter with the module replaced by an over-approximate summary, while relational strengthening proves the original information-flow property conjoined with an auxiliary cross-copy relation on the concrete two-copy RTL.}
\end{figure}
\vspace{-1mm}



\begin{figure*}[t]
    \centering
    \includegraphics[width=1\linewidth]{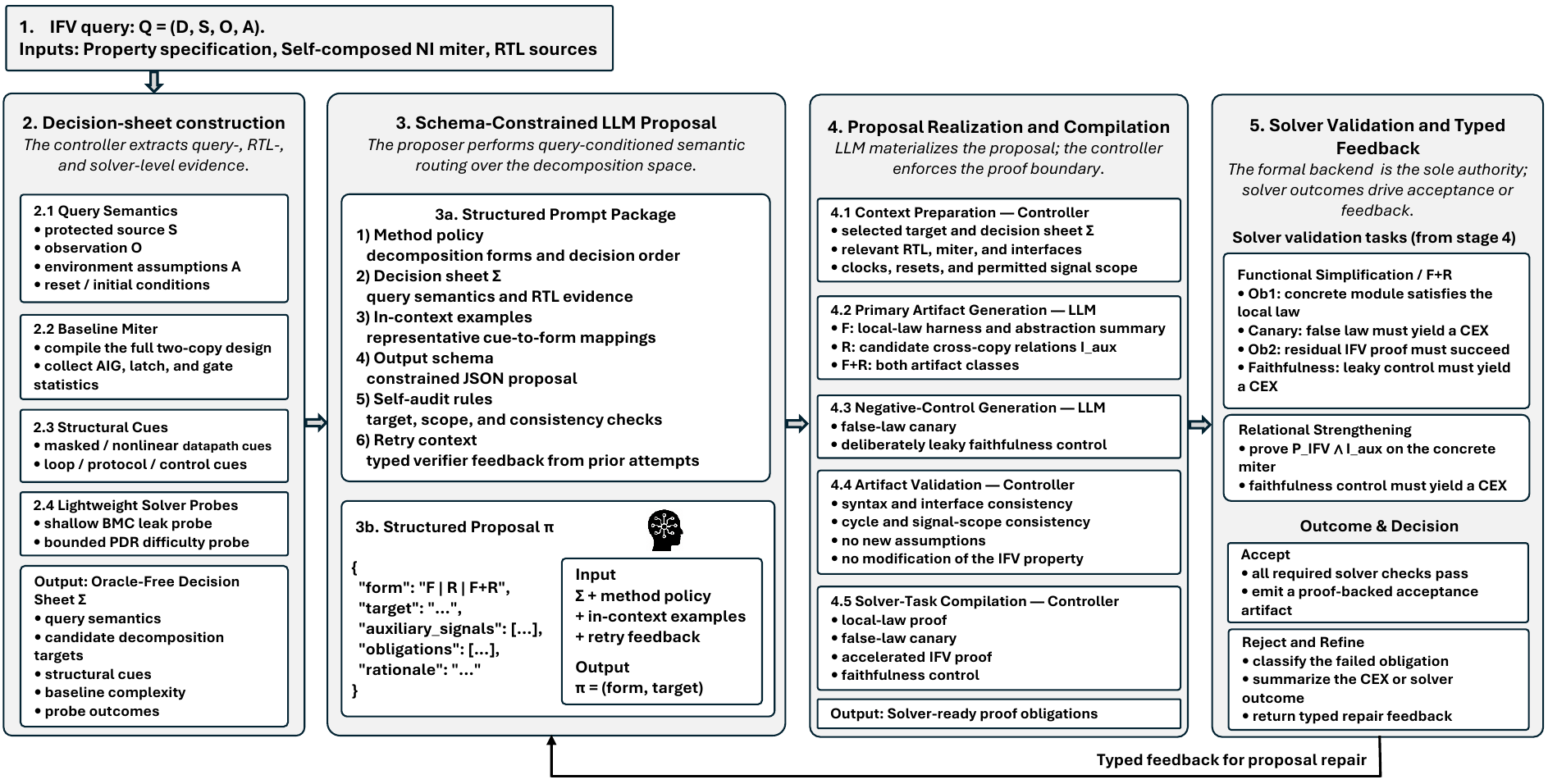}
    \caption{SLED-IFV closed-loop decomposition flow. The controller builds an oracle-free decision sheet \(\Sigma\); the LLM emits a strict JSON proposal \(\pi\) naming the decomposition form and target, then materializes the proposal into proof artifacts under controller checks. The controller compiles the validated artifacts into solver tasks, and the verifier accepts or rejects the candidate. Rejections are classified into typed feedback and returned for  repair.
    }
    \label{fig:framework}
    \Description{Flowchart of the SLED-IFV closed loop: a controller builds an oracle-free decision sheet from the query and RTL evidence; a large language model emits a JSON proposal naming a decomposition form and target and then materializes proof artifacts under controller checks; the controller compiles the validated artifacts into solver tasks; and property-directed reachability accepts the candidate or returns typed feedback for another attempt.}
\end{figure*}
\vspace{-2mm}

\section{Method: SLED-IFV}
\label{sec:method}
\vspace{-1mm}

\subsection{Overview}

SLED-IFV formulates RTL IFV acceleration as a solver-validated decomposition problem. Given an IFV query \(Q=(D,S,O,A)\), the controller first constructs the standard self-composed miter for \(P_{\mathsf{IFV}}\). The flow then searches for
\[
    \pi=(\mathsf{form},\mathsf{target}),
    \qquad
    \mathsf{form}\in\{\mathsf{F},\mathsf{R},\mathsf{F{+}R}\},
\]
where \(\mathsf{F}\) denotes functional simplification, \(\mathsf{R}\) relational strengthening, and \(\mathsf{F{+}R}\) their composition. The target is a module region \(M\), a set of public-control signals used to form \(I_{\mathsf{aux}}\), or both.

The method separates semantic proposal, proposal realization, and proof acceptance. The first LLM stage selects the decomposition form and target from an oracle-free decision sheet. A second LLM-guided realization stage then materializes the proposal into concrete verification artifacts, such as an obligation-1 (Ob1) harness, an abstraction stub, candidate \(I_{\mathsf{aux}}\), and negative controls. The controller constrains and checks these edits, compiles the well-formed artifacts into solver tasks, and invokes PDR. 
Thus, the LLM participates in both decomposition selection and artifact generation, but only the verifier determines whether a candidate is accepted.


\subsection{Semantic Proof Decomposition Forms}
\vspace{-1mm}
The two decomposition forms correspond to the proof barriers identified in Section~\ref{sec:motivation}. As shown in Fig.~\ref{fig:decomposition-forms}, functional simplification removes proof-hard implementation detail behind a validated boundary summary, whereas relational strengthening exposes cross-copy facts that PDR needs for induction.

\subsubsection{Functional simplification}
Functional simplification applies when a proof-hard RTL region lies in the sink cone, but the IFV argument depends only on a compact relation at its boundary. For a proposed region \(M\), the realization stage generates a contract \(C\), an Ob1 harness over the concrete module, and a summary \(M^{\#}\) that enforces \(C\) over selected boundary signals. Signals not constrained by \(C\) remain nondeterministic, so \(M^{\#}\) over-approximates the concrete implementation.

The controller compiles
\[
    \mathrm{Ob1}: M \models C,
    \qquad
    \mathrm{Ob2}: D[M^{\#}] \models P_{\mathsf{IFV}} .
\]
Ob1 proves that the concrete RTL satisfies the proposed boundary law. Ob2 proves the residual IFV property after replacing \(M\) with \(M^{\#}\). If both obligations hold, every concrete behavior of \(D[M]\) is represented by \(D[M^{\#}]\), and the original IFV claim follows.

Because \(M^{\#}\) is an over-approximation, an Ob2 counterexample need not correspond to a concrete leak. Unconstrained summary signals may admit boundary behaviors that satisfy \(C\) but cannot arise in \(M\). Such failures are returned as refinement feedback rather than reported immediately as violations. The next attempt may revise the contract, change the module boundary, add a relational component, or abandon functional simplification.

The realization stage also generates a false-law canary. This negative control replaces \(C\) with a deliberately incorrect law that must produce a counterexample. Failure to do so indicates a vacuous or malformed Ob1 harness.

\subsubsection{Relational strengthening}

Relational strengthening applies when the missing proof fact is a relation between the two executions rather than a local datapath summary. From the selected public-control or interface signals, the realization stage generates a candidate auxiliary relation \(I_{\mathsf{aux}}\), such as counter equality, protocol-phase alignment, valid-signal lockstep, or control-state equality.

The verifier must prove
\[
    D^{(0)} \parallel D^{(1)}
    \models
    P_{\mathsf{IFV}} \wedge I_{\mathsf{aux}} .
\]
Because \(P_{\mathsf{IFV}}\) remains a conjunct of the proved property, \(I_{\mathsf{aux}}\) is established rather than assumed. The controller restricts edits to the permitted signal set, rejects direct restatements of the sink equality, and disallows changes to the original property or assumptions.

\subsubsection{Composed decomposition}
When both barriers are present, the realization stage generates both the functional artifacts and the relational candidate. The controller combines the summary and strengthened miter, and the resulting candidate must pass the local contract proof, false-law canary, accelerated proof, and faithfulness control.

\vspace{-2mm}
\subsection{Closed-Loop Flow}
\vspace{-1mm}
Figure~\ref{fig:framework} shows the complete SLED-IFV flow, comprising decision-sheet construction, schema-constrained LLM proposal, LLM-guided artifact realization under controller validation, and solver-based acceptance or typed feedback for iterative repair.

\subsubsection{Decision-sheet construction}
The controller records \(S\), \(O\), and \(A\), constructs the baseline two-copy miter, and emits its AIG. Event-based observations are guarded when needed so that equality is checked only when the observation is valid. 
The oracle-free decision sheet \(\Sigma\) combines the source--sink--assumption triple, candidate modules and \(I_{\mathsf{aux}}\) signals, AIG statistics, RTL-text flags, and lightweight solver probes. Shallow BMC identifies bounded leaks, while short PDR runs identify already-easy cases. These probes guide proposal selection but do not contribute proof evidence. The sheet contains neither the expected decomposition nor the final proof outcome.


\subsubsection{Schema-Constrained LLM proposal}

The proposer receives a structured prompt package comprising the decision sheet \(\Sigma\), relevant RTL sources, and a method prompt that specifies the decomposition forms and decision order, in-context examples that illustrate representative mappings from source--sink patterns to functional, relational, or composed decompositions, and a strict JSON schema for the output proposal. Self-audit rules require the model to check the proposed target, permitted signal scope, and consistency with the supplied candidate lists before returning a result. On subsequent attempts, the failure type, solver outcome, and relevant counterexample summary from the previous attempt are added to the prompt, allowing the model to revise the rejected proposal rather than regenerate it without feedback. The resulting proposal specifies the decomposition form, concrete target, intended proof obligations, and a concise rationale linking the query and RTL features to the selected strategy. At this stage, the proposal identifies what should be decomposed and where the decomposition should be applied; the corresponding solver-ready artifacts are generated in the subsequent realization stage.

\subsubsection{Proposal realization and compilation}
The realization stage converts the semantic proposal into concrete proof artifacts. The controller first prepares the relevant context: the proposed target, the original miter, interfaces, clocks, resets, and permitted signal scope. Guided by this context, the LLM generates the primary artifacts required by the selected form. For functional simplification, these include the Ob1 harness and abstraction stub; for relational strengthening, they include candidate \(I_{\mathsf{aux}}\); for the composed form, both sets are generated. The LLM also produces the negative controls used by the validation flow, including the false-law canary and deliberately leaky faithfulness miter.
The controller then validates the generated artifacts before compilation. It checks syntax and interface consistency, enforces cycle alignment and the permitted edit scope, and rejects any new assumptions or changes to the original IFV property. Artifacts that fail these checks are returned as apply-time feedback. Valid artifacts are compiled into Ob1, false-law, accelerated-proof, and faithfulness jobs, together with the corresponding solver-ready AIGs and scripts.

\begin{table}[t]
\centering
\caption{Typed validation failures and corresponding repair directions.}
\label{tab:failure-feedback}
\scriptsize
\setlength{\tabcolsep}{3pt}
\renewcommand{\arraystretch}{1.0}
\begin{tabular}{@{}p{0.20\columnwidth}
                p{0.34\columnwidth}
                p{0.36\columnwidth}@{}}
\toprule
Failure & Meaning & Repair direction \\
\midrule
Apply rejected &
The artifact cannot be compiled or applied. &
Fix the syntax, interface, target, or build configuration. \\

Ob1 failed &
The proposed local law is false. &
Revise the law or module boundary. \\

Ob1 vacuous &
The false-law canary does not produce a CEX. &
Repair the Ob1 harness or negative control. \\

Ob2 failed &
The residual or strengthened IFV proof fails. &
Refine the summary, target, decomposition form, or
\(I_{\mathsf{aux}}\). \\

Faithfulness failed &
The deliberately leaky control does not produce a CEX. &
Inspect the property, assumptions, or abstraction. \\
\bottomrule
\end{tabular}
\end{table}
\vspace{-3pt}

\subsubsection{Solver validation and feedback}

Solver validation provides the final acceptance criterion. A functional candidate is accepted only when both Ob1 and Ob2 are proved, while the false-law canary and faithfulness control each produce a counterexample. A relational candidate must prove \(P_{\mathsf{IFV}} \wedge I_{\mathsf{aux}}\) on the concrete two-copy miter, with the faithfulness control still exposing the seeded leak. A composed candidate must satisfy the obligations of both forms.
If an obligation fails, the controller records the rejected artifacts, the solver outcome, and any relevant counterexample, and maps the failure to one of the categories in Table~\ref{tab:failure-feedback}. The resulting feedback is returned to the proposal and realization stages to guide the next attempt. For example, an Ob1 failure points to an invalid local law or module boundary, whereas an Ob2 failure following a successful Ob1 often indicates that the summary is too weak and admits spurious boundary behaviors. The loop continues until a candidate is accepted, a concrete violation is established, or the retry budget is exhausted.

\vspace{-3mm}
\subsection{Soundness Boundary}

The LLM influences decomposition search and artifact construction, but not proof acceptance. Functional simplification is accepted only after Ob1 proves the concrete contract and Ob2 proves \(P_{\mathsf{IFV}}\) on an over-approximation. Relational strengthening is accepted only after the concrete two-copy miter proves \(P_{\mathsf{IFV}}\wedge I_{\mathsf{aux}}\). Controller-side scope checks prevent generated artifacts from changing the original property or adding assumptions, while false-law canaries and faithfulness controls guard against the targeted vacuity and harness failures. Consequently, an incorrect proposal or generated artifact can consume time or trigger another iteration, but it cannot produce an accepted proof without satisfying the solver-checked obligations.


\begin{table}[t]
\centering
\caption{Solver validation runtime for accepted candidates. Ob1 is the local-law proof, Ob2 is the accelerated proof, and Faith reports the counterexample frame of the leaky control}
\label{tab:runtime}
\footnotesize
\setlength{\tabcolsep}{4pt}
\renewcommand{\arraystretch}{1.1}
\resizebox{\columnwidth}{!}{%
\begin{tabular}{llrrrrl}
\toprule
Case & Form & Baseline & Ob1 & Ob2 & Faith & Outcome \\
\midrule
\texttt{masked\_mult}      & F   & 39.3~s      & 0.04~s & 0.8~s  & CEX@4  & $\approx 47\times$ \\
\texttt{gs\_div}      & F   & 281.4~s      & 6.6~s & 0.46~s  & CEX@4  & $\approx 40\times$ \\
\texttt{otbn\_sec\_add}    & F   & 1224~s      & 0.03~s & 2.0~s  & CEX@9  & $\approx 603\times$ \\
\texttt{keccak\_chi}       & F   & 1459~s      & 2.4~s  & 56.6~s & CEX@5  & $\approx 25\times$ \\
\texttt{picorv32}          & R   & 212~s   & --     & 57.7~s & CEX@20 & $\approx 3.7\times$ \\
\texttt{ibex}              & R   & TO@12h      & --     & 55.8~s & CEX@13 & timeout-to-proof \\
\texttt{hmac}              & R   & 502~s       & --     & 21.7~s & CEX@87 & $\approx 23\times$ \\
\texttt{hmac\_shadone}     & R   & 8020~s      & --     & 26.0~s & CEX@87 & $\approx 310\times$ \\
\texttt{mini\_cipher}      & F+R & TO@12h      & 0.8~s  & 13.0~s & CEX@7  & timeout-to-proof \\
\bottomrule
\end{tabular}}
\end{table}
\vspace{-1mm}
\begin{table}[t]
\centering
\caption{Technique ablation on \texttt{mini\_cipher}. Only the combined
functional and relational decomposition proves within the 12-hour budget.}
\label{tab:tech-ablation}
\normalsize
\setlength{\tabcolsep}{12pt}
\renewcommand{\arraystretch}{0.750}
\resizebox{\columnwidth}{!}{%
\begin{tabular}{lllr}
\toprule
Config. & S-box & Round control & Result \\
\midrule
Baseline    & real masked & local    & TO@12h \\
Simpl. only & stub        & local    & TO@12h \\
Rel. only   & real masked & lockstep & TO@12h \\
Both        & stub        & lockstep & 13.0~s \\
\bottomrule
\end{tabular}}
\end{table}











\begin{table}[t]
\centering
\caption{Component ablation. B/V/R denote the numbers of benchmarks,
variants, and independent proposals. W/O and W/ indicate without and with each component.}
\label{tab:component-ablation}

\resizebox{\columnwidth}{!}{%
\begin{tabular}{@{}l l c l c c@{}}
\toprule
\textbf{Component} &
\textbf{Task} &
\textbf{B/V/R} &
\textbf{Metric} &
\textbf{W/O} &
\textbf{W/} \\
\midrule
Candidate lists
& Target selection
& \(10^{a}/1/3\)
& Correct target
& 13/30
& 24/30 \\

Cheap probes
& Leak/easy triage
& \(5^{b}/2/3\)
& Correct routing
& 6/30
& 30/30 \\

Verifier feedback
& End-to-end repair
& \(9^{c}/1/1\)
& Accepted proof
& 6/9
& 9/9 \\

Few-shot examples
& Form selection
& \(10^{a}/1/3\)
& Correct form
& 22/30
& 30/30 \\

RTL-text flags
& Target selection
& \(10^{a}/1/3\)
& Correct target
& 19/30
& 24/30 \\
\bottomrule
\end{tabular}%
}
\vspace{-1mm}

\vspace{0.35em}
\begin{minipage}{\columnwidth}
\scriptsize
\raggedright
\(^{a}\) Nine nontrivial proof cases plus the already-easy
\texttt{chipyard\_muldiv} routing control.
\quad
\(^{b}\) Four deliberately leaking cases plus the already-easy routing control,
each evaluated with and without probe evidence.
\quad
\(^{c}\) Nine nontrivial proof cases used for end-to-end validation.
\end{minipage}
\vspace{-1mm}
\end{table}

\vspace{-1mm}
\section{Evaluation}
\label{sec:evaluation}
\vspace{-1mm}

We evaluate three aspects of SLED-IFV: whether the proposed decompositions accelerate hard IFV proofs under a fixed backend, whether functional simplification and relational strengthening address distinct proof barriers, and which components of the proposal loop are necessary for reliable automation.

\vspace{-2mm}
\subsection{Setup and Benchmarks}
\label{subsec:eval-setup}
\vspace{-1mm}

We construct nine nontrivial self-composed RTL IFV benchmarks from public designs and published masking components. Four cases \texttt{masked\_mult}, \texttt{gs\_div}, \texttt{otbn\_sec\_add}, and \texttt{keccak\_chi} exercise functional simplification on proof-hard datapaths, including masked cryptographic logic and iterative floating-point arithmetic. Four cases \texttt{picorv32}, \texttt{ibex}, \texttt{hmac}, and \texttt{hmac\_shadone} exercise relational strengthening on control or timing proofs. The remaining case, \texttt{mini\_cipher}, requires both forms. We additionally use \texttt{chipyard\_muldiv} as an already-easy routing control, which is identified as trivial by the preprocessing probes and therefore bypasses decomposition.

For each benchmark, we define the protected source, public observation, environment assumptions, and target non-interference property. The underlying RTL includes OpenTitan-derived masked datapaths and HMAC/SHA logic, a Goldschmidt floating-point divider, lowRISC Ibex, YosysHQ PicoRV32, and a Chipyard/BOOM-derived control block. Two miters reuse OpenTitan DOM masking gadgets rather than synthetic substitutes. All benchmarks share public inputs and assumptions across the two copies, allow independent protected sources, and check equality at the selected public observation.

Experiments run on an Intel Core i9-14900K workstation with 188~GiB RAM and Ubuntu~22.04.5. We use Yosys~0.65~\cite{Yosys,yosys-paper} as the RTL front-end and \texttt{sv2v}~0.0.13 when SystemVerilog conversion is required. Verification uses UC Berkeley ABC~1.01~\cite{abc}: proof obligations run with \texttt{pdr}, while bounded probes and faithfulness controls use \texttt{bmc3}. All proofs are single-threaded, and 12-hour baselines run in isolation. 
Claude Opus~4.7 (\texttt{claude-opus-4-7}), invoked through Claude Code CLI~2.1.141, serves as both the proposer and artifact generator. Claude Sonnet~4.6 (\texttt{claude-sonnet-4-6}) and Claude Haiku~4.5 (\texttt{claude-haiku-4-5-20251001}) are used only for the weak-model ablations.

\vspace{-1mm}
\subsection{Proof Acceleration}
\label{subsec:eval-runtime}
\vspace{-1mm}

Table~\ref{tab:runtime} reports solver time along each accepted SLED-IFV path. For functional simplification, Ob1 proves the local law on the concrete module and Ob2 proves the residual IFV property after summary replacement. For relational strengthening, Ob2 proves \(P_{\mathsf{IFV}}\wedge I_{\mathsf{aux}}\) on the concrete two-copy RTL. In the composed case, Ob2 includes both transformations. The reported times exclude LLM latency.


SLED-IFV reduces minute-to-hour baselines to second-scale obligations and completes two cases that PDR cannot prove within 12 hours. For the functional cases, the LLM-guided realization replaces proof-hard datapath regions with solver-validated over-approximate summaries. For the masked cases, the summaries leave individual output shares nondeterministic while constraining their recombined values. For \texttt{gs\_div}, SLED-IFV instead summarizes each Goldschmidt iteration with a multiplication-free convergence contract that lower-bounds the growth of the denominator's leading-one prefix. For the relational cases, it strengthens the original two-copy miters with selected equalities over key-independent processor or HMAC/SHA control state. For \texttt{mini\_cipher}, it combines a summarized masked-GF datapath with round-control alignment. These transformations achieve up to \(603\times\) solver-only speedup, reduce all completed proofs to 0.84--59 seconds, and convert both \texttt{ibex} and \texttt{mini\_cipher} from 12-hour timeouts into completed proofs. All applicable false-law canaries and faithfulness controls produce counterexamples, confirming that the tested harnesses are non-vacuous and remain sensitive to seeded leaks.


Across the 9 nontrivial cases, one LLM pass, including proposal and artifact generation, takes approximately 25 seconds on average, only about 0.2\% of the average plain-PDR baseline runtime when 12-hour timeouts are counted at the cutoff. The flow typically converges within one to four passes.
We therefore report solver-only speedup in Table~\ref{tab:runtime} to isolate the reduction in proof difficulty.

\vspace{-2mm}
\subsection{Technique Complementarity}
\label{subsec:eval-tech-ablation}
\vspace{-1mm}

Table~\ref{tab:tech-ablation} isolates the two barriers in \texttt{mini\_cipher}, which combines a masked DOM S-box with an iterative round loop. Replacing only the S-box removes the masked-datapath cost but leaves the round-alignment proof unresolved. Adding only round-control lockstep exposes the required relational invariant but leaves the nonlinear masked implementation intact. Both single-form configurations time out, whereas their composition proves in 13.0 seconds. The two decompositions therefore expose complementary proof facts rather than interchangeable solver hints.

\vspace{-2mm}
\subsection{Component Ablation}
\label{subsec:eval-components}
\vspace{-1mm}

Table~\ref{tab:component-ablation} evaluates each automation component in the setting for which it was designed: candidate lists and RTL-text flags for target selection, cheap probes for leak and trivial-case triage, verifier feedback for target repair, and few-shot examples for mapping observed cues to decomposition forms. All ablations use Claude Sonnet, except the few-shot study, which uses Claude Haiku to make prompt-example effects more visible under a weaker proposer.

Candidate lists improve target selection from \(13/30\) to \(24/30\) by exposing the concrete module and signal names. Cheap probes serve a different purpose. On leaking and already-easy variants, they raise correct routing from \(6/30\) to \(30/30\), preventing unnecessary decomposition attempts before solver time is spent.
Verifier feedback closes the gap between selecting the right form and producing a verifier-accepted candidate. The one-shot proposer chooses the correct form in all nine runtime cases, but three proposals identify the wrong target. Typed feedback repairs all three, increasing acceptance from \(6/9\) to \(9/9\). Few-shot examples improve form selection from \(22/30\) to \(30/30\) under the weaker model, while RTL-text flags provide a smaller target-selection gain from \(19/30\) to \(24/30\).

Overall, the components play distinct roles. Candidate lists ground concrete targets, probes handle leak and trivial-case triage, verifier feedback repairs failed proposals, and prompt-side examples and RTL cues support decomposition selection.



\vspace{-2mm}
\section{Conclusion}
\label{sec:conclusion}
\vspace{-1mm}


We presented SLED-IFV, a solver-validated LLM-guided decomposition framework for scalable hardware information-flow verification. SLED-IFV uses the LLM to select query-specific decomposition forms and targets and to materialize the corresponding proof artifacts, while the formal backend remains the sole authority for acceptance. Across nine nontrivial RTL IFV cases, SLED-IFV reduces minute-to-hour PDR proofs to second-scale obligations, achieves up to \(603\times\) solver-only speedup, and converts two 12-hour timeouts into completed proofs. The results further show that functional simplification and relational strengthening address distinct proof bottlenecks and can be composed when both arise in the same query. More broadly, SLED-IFV demonstrates that LLMs can assist formal hardware verification by automating semantic proof decomposition without expanding the trusted computing base.

\begin{acks}
This work was partially supported by a grant from the CHEST IUCRC.
\end{acks}

\bibliographystyle{ACM-Reference-Format}
\bibliography{refs}

\end{document}